\documentclass[twocolumn,amsmath,amssymb,graphicx]{revtex4-1}

\draft 

\usepackage{graphicx} 
\usepackage{dcolumn} 
\usepackage{bm} 
\usepackage[version=3]{mhchem}

\begin{document}



\title{Integration and characterization of solid wall electrodes in microfluidic 
devices fabricated in a single photolithography step}

\author{T.~W.~Herling}

\author{T.~M\"uller}

\author{L.~Rajah}
\affiliation{Department of Chemistry, University of Cambridge, Lensfield Road, Cambridge CB2 1EW, United Kingdom}

\author{J.~N.~Skepper}
\affiliation{Department of Physiology, Development and Neuroscience, University of Cambridge, Downing Street, Cambridge CB2 3DY, United Kingdom}

\author{M.~Vendruscolo}
\affiliation{Department of Chemistry, University of Cambridge,  Lensfield Road, Cambridge CB2 1EW, United Kingdom}

\author{T.~P.~J.~Knowles}
\email{tpjk2@cam.ac.uk}
\affiliation{Department of Chemistry, University of Cambridge,  Lensfield Road, Cambridge CB2 1EW, United Kingdom}

\date{\today}

\begin{abstract}
We describe the fabrication and characterization of solid 3-dimensional electrodes in 
direct contact with microfluidic channels, implemented using a single 
photolithography step, allowing operation in high-dielectric constant media. Incorporation and self-alignment of electrodes is 
achieved by combining microsolidic approaches with exploitation of the surface 
tension of low melting point alloys. Thus the metal forms the walls flanking the channel. We show that this approach yields 
electrodes with a well-defined, reproducible morphology and stable electronic 
properties when in contact with biochemical buffers. By combining calibration of the electric field with free-flow electrophoresis we quantify the net solvated charges of small molecules.
\end{abstract}


\maketitle

\begin{figure}[h!] 
\centering
\includegraphics[width=84mm]{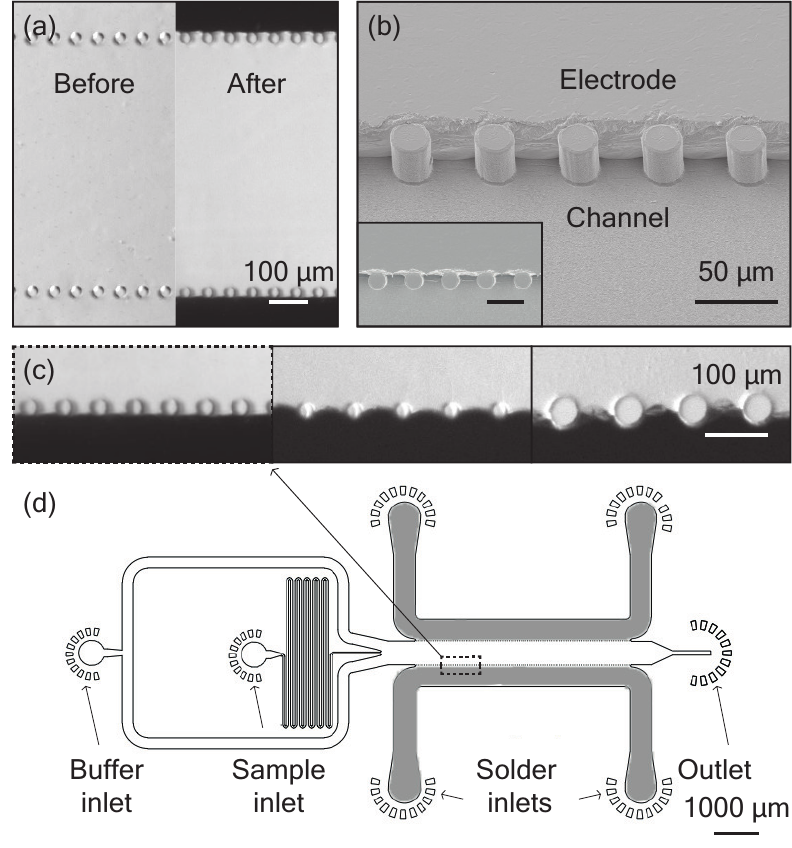}
\caption[design]{ (a) Brightfield image 
	of the main sample channel before (left) and after (right) electrode 
	incorporation. The channel is flanked by PDMS pillars preventing the 
	solder (opaque) from entering the main channel during device fabrication. (b) Scanning electron micrograph (SEM) of the 
	electrode-channel interface at a $40^\circ$ tilt from the position of the 
	glass, below the device. SEM image (insert) of solder and pillars at the 
	channel edge, after the removal of a device from its glass substrate. (c) Variation in pillar spacing and diameter. From left to right; 25~$\mu$m pillars 25~$\mu$m apart, 25~$\mu$m pillars 50~$\mu$m apart, and 50~$\mu$m pillars and spacing. (d) 
	The device design used for characterization of the small molecules through. Solder, 
	sample and buffer inlets are indicated, as is the liquid outlet. Electrode 
channels have been shaded in gray.}
\label{fig:design}
\end{figure}

The ability to control electric fields within microfluidic devices is the 
foundation for many key applications of this technology\cite{Voldman2002, 
Burg2003, Ahn2006, Burg2007, Kohlheyer2008, He2009, Abate2010, Agresti2010, 
Mojarad2012, ODonovan2012, Song2013}. Approaches exist to use microsolidics 
techniques\cite{Siegel2007} in combination with soft lithography to generate 
electrodes within separate channels that are suitable for use in low-dielectric media such as oils utilized in digital 
microfluidics\cite{Abate2010}; however, strong electrostatic screening in 
conductive aqueous media from ionic species present in solution necessitates the 
use of fundamentally different approaches to generate strong fields in such 
media.  Indeed, direct electric contact between the electrodes and the 
conductive sample media, such as aqueous buffer solutions, is required to 
prevent screening at the electrode interface.  This requirement presents a 
challenge for micro-fabrication, and conventional methods for the integration of such elements rely on multistep 
processes and controlled alignment of such elements to ensure the accurate relative 
positioning of the fluidic and electronic components of the 
device\cite{Kohlheyer2006, Kim2007, Ebina2009, Romanowsky2010, Koehler2011, Cheng2011}. 
 In this Letter, we combine microsolidics techniques with exploitation of the 
surface tension of molten InBiSn alloy to fabricate 3-dimensional (3D) electrodes in direct 
contact with aqueous media within microfluidic devices\cite{So2011, Li2012}. We 
characterize the electronic properties of the devices and show that this 
approach allows a stable electrode/solvent interface to be positioned with micrometer precision as shown in  Fig.~\ref{fig:design}(a) to (c). Furthermore, we demonstrate the use of 
such electrodes in free-flow micro-electrophoresis to determine accurately the charges 
of molecular species in solution.

In order to create a direct contact between the metal electrodes and electrolyte within microfluidic 
channels, we generated, using soft-lithography, arrays of pillars to define the 
position of the electrodes within the channels\cite{So2011}. The corresponding sections at the sides of the channel were subsequently filled with a molten InBiSn alloy which was allowed to 
solidify within the device. Resulting in a large 
active area and a vertically homogeneous electric field. This is in contrast to conventional approaches where the electrodes are 
patterned onto the basis of the microchannels.


The microchannels were fabricated in polydimethylsiloxane (PDMS) on glass substrates using standard soft lithography techniques\cite{McDonald2002}. In order to minimize 
experimental noise, opaque devices were produced 
by mixing a small fraction of carbon nano powder, 0.2~\%~w/w, into the PDMS prior 
to curing.  Electrodes, see Fig.~\ref{fig:design}(a) and (c), were 
incorporated by inserting a low melting point InBiSn alloy (51\%~In, 32.5\%~Bi, 16.5\%~Sn, Conro Electronics) through the 
designated channels of devices placed, glass slide down, on a hot plate set to 
$79^{\circ}\rm{C}$. At this temperature, the alloy melts upon contact with the 
glass substrate and fills the designated channels upon light pressure, but is 
sufficiently viscous and possesses a sufficient surface tension so as not to 
flow between the pillars. The inserted electrodes then solidify at room 
temperature. The fabrication process therefore enables the production of 
microfluidic devices incorporating 3D electrodes automatically aligned with 
fluidic components and in direct contact with them.

This approach allows pillar arrays with a range of dimensions to be used in 
conjunction with the surface tension of molten solder to integrate electrodes in 
microfluidic devices, as demonstrated in Fig.~\ref{fig:design}(c).  An upper 
limit to pillar size and spacing is given by the requirement to prevent the flow 
of the molten metal between the pillars; in practise effective containment is 
lost for spacings exceeding 75~$\mu$m. The practical lower limit to pillar 
sizing is given by the soft lithography process used to define them\cite{McDonald2002}
In the remainder of this Letter, we work with a pillar diameter and spacing of 
25~$\mu$m in each case.


\begin{figure}[th] 
\centering
\includegraphics[width=80mm]{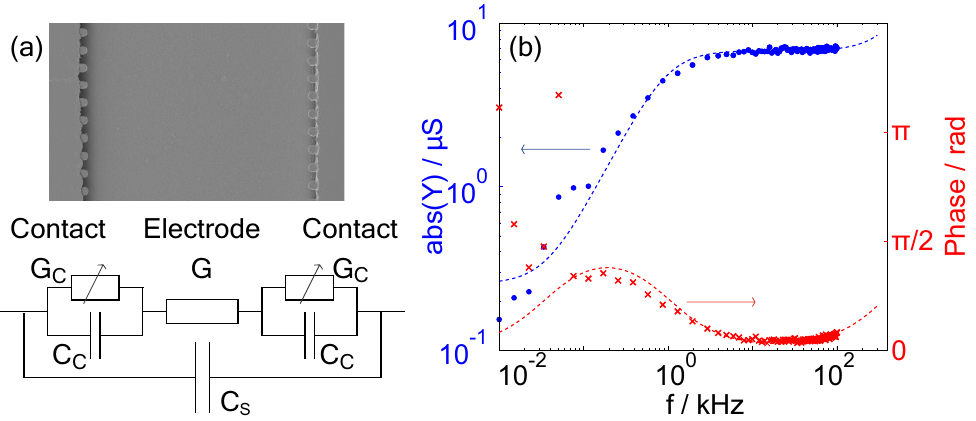}
\caption[conduct]{(a) Top: SEM image of the electrodes and channel taken from 
	the bottom of the channels which are eventually sealed with a glass slide.  
	Bottom: Diagram of the equivalent circuit model of our devices.
The conductive channel is represented by a conductor $G$ which is attached
to the electrodes via two contacts characterized by a voltage-dependent
conductivity $G_C$ and a contact capacitance $C_C$. The entire device has an
additional (stray) capacitance $C_S$ in parallel. (b) Measurement of absolute value (blue dots, left axis) and phase (red x-es, right axis) of the admittance $Y$ of a sample device filled with
conductivity standard ($500~\mu\textrm{S/cm}$) as a function of frequency. The dashed lines are fits to the complex data using the circuit model depicted in (a) with $G$, $G_C$, $C_C$ and $C_S$ as free parameters. }
\label{fig:conduct}
\end{figure}

We characterized the stability of the electronic properties of the 3D electrodes 
in solution and their suitability for applying controlled electric fields in 
aqueous media within microfluidic devices by repeatedly measuring the 
frequency-dependent admittance of the device shown in Fig.~\ref{fig:design}(c) 
and comparing the results with a suitable circuit model.  A scheme of an 
equivalent circuit for the device is presented in Fig.~\ref{fig:conduct}(a), where 
the fluid channel is modeled as a conductance $G$ in series between two 
solder-to-fluid contacts, characterized by a (low) conductance $G_C$ and a 
parallel capacitance $C_C$, with an overall stray capacitance $C_S$. To perform 
the calibration measurement, a $10~\textrm{mV}$ AC voltage with variable 
frequency was applied over a device filled with a $500~\mu\textrm{S/cm}$ 
conductivity standard. The resulting current, and thus the admittance $Y$, was 
then measured as a voltage drop over a $220~\Omega$ resistor using a lock-in 
amplifier and is shown in Fig.~\ref{fig:conduct}(b).

At very low frequencies, the overall conductance is limited by the low 
conductance of the contacts. At intermediate frequencies of approximately 
$100~\textrm{Hz}$, the conductance through the contact capacitance exceeds the 
one through the contact resistance, leading to a linear increase of the admittance with the frequency. Above 
$1~\textrm{kHz}$, the conductance of the channel itself becomes the limiting 
factor governing the overall current flow, and the absolute value of the 
admittance is constant with frequency until conductance through the stray 
capacitance becomes appreciable. Comparison to the modeled, complex 
frequency-dependent admittance (dashed lines in Fig.~\ref{fig:conduct}(b) 
allows for determination of the circuit parameters introduced above, and we obtain 
contact capacitances of the order of a few nF, stray capacitances in the range 
of a few pF and contact resistances much larger than $1~\textrm{M}\Omega$.  
Moreover, using the given conductivity of the standard solution thus allows us 
also to determine the cell constant $K = \sigma/G$ linking 
conductance $G$ and conductivity $\sigma$ in each device.

Our model fits the experimental data very well, with noticeable deviations only at very low frequencies where conductance 
is low.  Furthermore, since the value of $G$ is directly given by the constant part of the spectrum and is independent of the values of the other fit parameters. The values we obtain for the individual cell constants are highly reliable and reproducible between devices. The mean cell constant and standard deviation for 10 devices were found to be $41\pm2.6~\rm{cm}^{-1}$. These results also agree well with calculated values\cite{notecalc}, but variation outside the overlap of error bounds were observed when investigating different devices in spite of the shared lithographic 
dimensions, highlighting the advantage of the direct experimental calibration afforded by this setup.

The functionality of these well-characterized electrodes can be explored by 
performing quantitative free-flow micro electrophoresis. A key requirement for charge quantification of molecular species by 
electrophoresis in free solution is the ability to determine accurately both the flow profile and 
the electric field strength across the solution where sample deflection occurs.
In order to quantify the electric field inside the flow
channel accurately, we use the calibrated cell constants $K$ for the individual 
devices and measure the conductivities of each of the buffers used. Recording of the DC
current flowing through a given device upon application of a voltage
then permits the effective voltage drop over the liquid to be
determined via $V=I/G$, which in turn yields the electric field
through $E=V/w$, with $w=663\pm33~\mu\textrm{m}$ the effective width of
the channel\cite{notewidth}.

To study the electrophoretic mobility of analytes under controlled conditions of 
pH and ionic strength, simultaneous electric current and analyte deflection 
measurements were undertaken for a range of voltages in three common biochemical 
buffer systems.  Measurements were made in HEPES, Tris and phosphate buffers, 
each at concentrations of 2.5, 5 and 10 mM, and all at pH 8.0. The fluorescent 
dyes fluorescein and rhodamine 6G were chosen as a model analytes.

\begin{figure}[th]
\centering
\includegraphics[width=80mm]{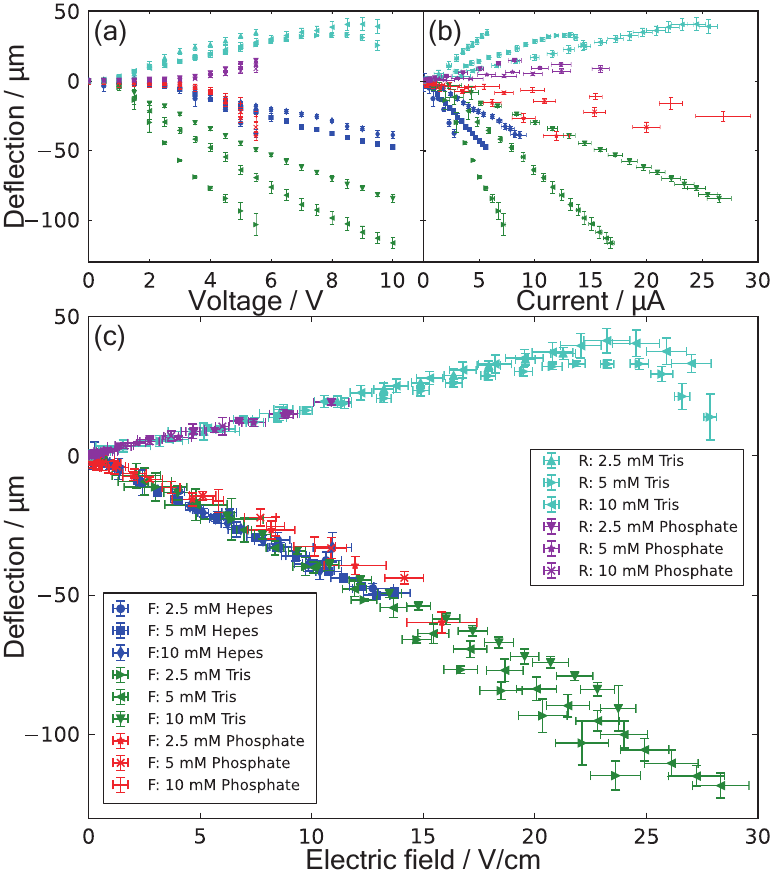}
\caption[dv]{(a) Analyte deflection $\delta$ measured at the
end of the 5 mm electrode main channel as a function of applied
voltage for different buffers. (b) The same deflection data set as in (a) but
shown versus the simultaneously measured current through the device.
(c) The same deflection measurements as a function of electric field
strength $E$ as determined via measured current and calibrated
conductance. All vertical error bars represent the statistical
spread of averaging the graphical deflection data over 25 pixels in
flow direction for 4 repeats. The horizontal error bars are the
standard deviation of 4 different repeats (b) and a combination
thereof with the uncertainty of the conductivity-to-conductance
calibration or calculation (c). The legend in (c) refers to all data shown in this figure. Entries preceded by F and R are for data with fluorescein and rhodamine 6G respectively.} 
\label{fig:dv}
\end{figure}

Experiments were performed at a flow rate of $250\pm0.88~\mu\textrm{L/h}$. The 
flow velocity in the central $450~\mu\textrm{m}$ wide section of the channel, 
averaged in the vertical direction, is to a good approximation constant with 
lateral position across the channel due to the high aspect ratio. We evaluated 
the average flow velocity in this areato be 
$4.97\pm0.5\times10^{-3}~\rm{m/s}$, resulting in a mean residence time between 
the electrodes of $0.99\pm0.1~\textrm{s}$ .


The position $\delta$ of the fluorescence intensity maximum of a
flow-focused fluorescein or rhodamine beam in different buffer solutions upon 
application of a
DC voltage to the electrodes is shown in Fig.~\ref{fig:dv}(a).
Data were recorded using a CCD camera operating through long working distance 
inverted optics with an integration time of
$1~\textrm{s}$ and averaged over 25 pixels corresponding to
$80~\mu\textrm{m}$ along the flow direction, with four
repeats.

Sample deflection of the order of tens of $\mu\textrm{m}$ was observed 
in all buffer systems investigated. However the observed voltage threshold below 
which ionic screening at the electrode-liquid interface inhibits the current 
between them, and thereby analyte deflection, varies between the three buffers. Furthermore the voltage drop across the solution relative to that 
at the metal-liquid interface varies with buffer concentration, affecting the 
extent of analyte deflection. These observations emphasize that the electric 
field in the solution is not simply proportional to the overall voltage drop, 
and highlights the requirement for simultaneous current measurements to 
calculate the electric field strength from the effective voltage drop over the 
solution.

When moving close to the electrodes, reaction of the dyes with electrolysis 
products may result in loss of fluorescence, as seen from the apparent decrease 
in the deflection data for rhodamine 6G in 5 and 10 mM Tris at higher applied 
voltages in Fig.~\ref{fig:dv}. This observation does not 
affect the dye at locations further from the electrodes, defining the 
dynamic operating range of the setup.

A near-linear dependence of analyte deflection on current is observed when the 
same deflection data is plotted against the simultaneously measured current in 
Fig.~\ref{fig:dv}(b). However, due to the differences in buffer conductivities a 
range of deflections are seen for equal currents. We note that at these low 
voltages and currents the nucleation and bubble formation of electrolysis 
products\cite{Kohlheyer2006, Kim2007, Koehler2011} is avoided.

\begin{figure}[h] 
\centering
\includegraphics[width=80mm]{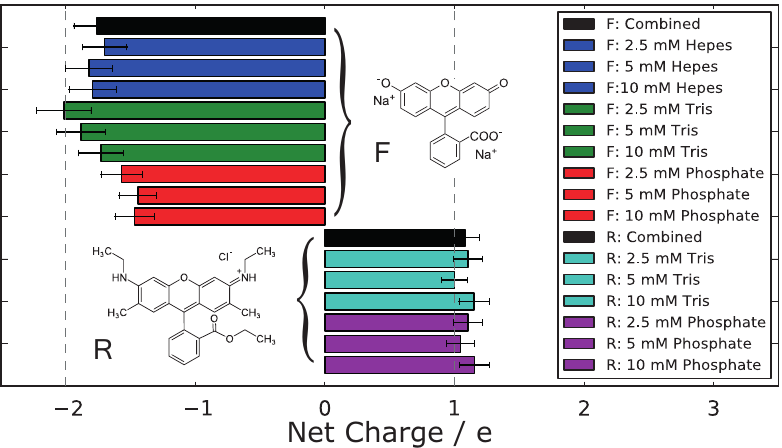}
\caption[bar]{Net charges determined for fluorescein and rhodamine 6G using 
the electrophoretic mobilities found from data for individual buffer systems and the combined 
mobilities for each dye. The label shows the buffer conditions and dye used, F 
for fluorescein and R for rhodamine 6G, and the chemical structures of the two dyes 
are shown. Dashed vertical lines represent the anticipated charges of $-2$ (fluorescein) 
and $+1~e$ (rhodamine 6G), respectively. }
\label{fig:bar}
\end{figure}

Remarkably, when the deflection is plotted as a function of calibrated
electric field in Fig.~\ref{fig:dv}(c), the data collapse accurately into an almost linear relationship, the slope of which defines the electrophoretic mobility $\mu_p=v_{p}/E$. To avoid perturbation of $\delta$ by ion depletion due to the low flow velocities at channel edges or reaction with electrolysis products, linear fits were performed to data for field strengths below $15~\rm{V/cm}$. In this manner, the electrophoretic mobility of the molecular 
species is found from division of the mean of these slopes by the residence time in the channel 
($0.99\pm0.1~\rm{s}$) to give electrophoretic mobilities of
$-3.72\pm0.36\times10^{-8}~\textrm{m}^2/\textrm{Vs}$ for fluorescein and 
$1.75\pm0.18\times10^{-8}~\textrm{m}^2/\textrm{Vs}$ in the case of rhodamine 6G. The 
main contribution to the error on the residence time is the height of the 
channel $25\pm2.5~\mu\rm{m}$. All devices used in the experiments presented here 
were fabricated using the same lithography master, hence variation is not 
expected between devices.

The measured electrophoretic mobilities can be converted to the corresponding solvated charge for molecules of known radii, here we used $4.5~\rm{\AA}$ for the radius of fluorescein\cite{Montermini2002} and $5.9~\rm{\AA}$ for rhodamine 6G\cite{Muller2008}. These values correspond to solvated charges of $-1.8\pm0.2~e$ for fluorescein and $+1.1\pm0.1~e$ for rhodamine 6G, see black bars in Fig.~\ref{fig:bar}. These charges determined under solution conditions are in good agreement with those anticipated from the dissociation of the dyes and counter ions shown in the insets of Fig.~\ref{fig:bar}.

In conclusion, an adaptable design and fabrication method for the incorporation 
of solid 3D electrodes of complex geometries into microfluidic devices is 
presented.  A single photolithography step is required in the fabrication of 
inherently aligned solid electrodes in direct contact with sample solutions.  
Free-flow electrophoresis was performed in a voltage range where electrolysis 
products are carried away by the liquid flow through the device. Three buffer 
systems were characterized at low mM concentrations. Simultaneous measurements were made 
of both analyte deflection and current at the application of increasing 
voltages. The cell constants of the devices were calibrated using conductivity 
standards. Conductivity measurements of the individual buffer solutions 
were used to quantify the electric field strength within the electrolyte, enabling the determination of the electrophoretic mobility and the charge of molecules in solution.

Support from the Biotechnology and Biological Sciences Research Council (BBSRC), 
the Frances and Augustus Newman Foundation, the Engineering and Physical Sciences Research Council 
(EPSRC) and the Swiss National Science Foundation (SNF) is gratefully acknowledged.


\begin{thebibliography}{25}
\bibitem [{not({\natexlab{a}})}]{notecalc}%
  \BibitemOpen
  \href@noop {} {} \ \bibinfo {note} {The calculation was
  performed by numerically iterating the Poisson equation with the potentials
  of rectangular electrodes $640~\mu\textrm{m}$ apart as boundary conditions
  and comparing the applied voltage to the resulting electric field evaluated
  across the entire length of the device (extending well beyond the end of the
  electrodes).}\BibitemShut {Stop}%
\bibitem [{not({\natexlab{b}})}]{notewidth}%
  \BibitemOpen
  \href@noop {} {} \ \bibinfo {note} {The measured width
  between the faces of the electrodes was $613\pm13~\mu\textrm{m}$. To account for the decreased channel area between the pillars their diameter of
  $25~\mu\textrm{m}$ was counted twice and a generous error estimate of $5\%$ attributed to the
  conversion from effective voltage drop to electric field.}\BibitemShut
  {Stop}%
\end{thebibliography}
\end{document}